\newcommand{\un}{~\mathrm}

\documentclass[namedreferences]{kluwer}

\usepackage{graphicx}
\usepackage{array}
\usepackage{dcolumn}
\usepackage{bm}
\usepackage{klucite}

\begin{document}
\begin{article}

\begin{opening}
\title{Anisotropic self-affine properties of experimental fracture surfaces}

\author{L. 
\surname{Ponson}$^{1,2}$\email{ponson@drecam.cea.fr}, 
D. \surname{Bonamy}$^1$, H. 
\surname{Auradou}$^2$, G. \surname{Mourot}$^3$, 
S. \surname{Morel}$^3$, E. 
\surname{Bouchaud}$^1$, C. \surname{Guillot}$^1$, 
J.P. \surname{Hulin}$^2$}
\institute{$^1$Fracture Group, Service de 
Physique et Chimie des Surfaces et Interfaces, 
DSM/DRECAM/SPCSI, B{\^a}timent 462, CEA Saclay, 
F-91191 Gif sur Yvette, France\\
$^2$Laboratoire Fluides, Automatique et Syst{\`e}mes 
Thermiques, UMR 7608, Universit{\'e} Paris 6 and 11, 
B{\^a}timent 502, Universit{\'e} Paris Sud, 91405 Orsay 
Cedex, France\\
$^3$Laboratoire de Rh{\'e}ologie du Bois de Bordeaux, 
UMR 5103, Domaine de l'Hermitage, 69 route 
d'Arcachon, 33612 Cestas Cedex, France}

\begin{abstract}
The scaling properties of post-mortem fracture
surfaces of brittle (silica glass), ductile (aluminum alloy) 
and quasi-brittle (mortar and wood) materials have been investigated. 
These surfaces, studied far from the initiation, were
shown to be self-affine. However, the Hurst exponent measured  
along the crack direction is found to be different from the one measured 
along the propagation direction. 
More generally, a complete description of the scaling properties of these 
surfaces call for the use of the 2D height-height correlation 
function  that involves three exponents $\zeta\simeq 0.75$, 
$\beta \simeq 0.6$ and $z \simeq 1.25$ independent of the material considered
as well as of the crack growth velocity. These exponents are shown to 
correspond to the roughness, growth and dynamic exponents 
respectively , as introduced in interface growth models.  
They are conjectured to be {\em universal}.
\end{abstract}


\abbreviations{\abbrev{KAP}{Kluwer Academic Publishers};
    \abbrev{compuscript}{Electronically submitted article}}

\nomenclature{\nomen{KAP}{Kluwer Academic Publishers};
    \nomen{compuscript}{Electronically submitted article}}

\classification{JEL codes}{D24, L60, 047}
\end{opening}

\section{Introduction}\label{intro}
Since the early work of \inlinecite{Mandelbrot}, 
the study of the morphology of fracture surfaces
is a very active field of research. Crack 
surfaces were observed
to display some universal scaling features 
although they result from a broad variety of 
specific processes taking place at the 
microstructure scale
(see \opencite{Bouchaud4}, for a review). Many 
experimental results suggested indeed that 
fracture surfaces are self-affine over a wide 
range of length scales In other words, the 
height-height correlation function $\Delta 
h(\Delta r)=<(h(r+\Delta r) - h(r))^2>^{1/2}_r$ 
computed along a given direction is found to 
scale as :

\begin{equation}
\frac{\Delta h}{l} = \left(\frac{\Delta r}{l}\right)^H
\end{equation}

\noindent where $H$ is  the Hurst exponent and 
$l$ the topothesy or the scale at which $\Delta 
h$ is equal to $\Delta r$. This exponent was 
found to be weakly dependent on the nature of the 
material and on the fracture mode with $H \simeq 
0.8$ (\opencite{Bouchaud9}; \opencite{Maloy}; 
\opencite{Schmittbuhl5}; \opencite{Daguier}). 
This exponent was then conjectured to be {\it 
universal} (\opencite{Bouchaud9}; 
\opencite{Maloy}).

Since the early 90s, a large amount of 
theoretical studies suggested scenarios to 
explain these experimental observations. They can 
be classified into two main approaches:
\begin{itemize}
\item[(i)]  Percolation-based models where the 
fracture propagation is assumed to result from a 
damage coalescence process (\opencite{Roux2}; 
\opencite{Hansen}). These models lead to 
isotropic fracture surfaces.
\item[(ii)] Elastic string models that modelize 
the crack front as an elastic line propagating in 
a random medium (\opencite{JPBouchaud}; 
\opencite{Ramanathan}; \opencite{Ramanathan2}). 
The fracture surface corresponds then to the 
trace left by this crack front. As emphasized by 
\inlinecite{Ponson5}, such approaches lead to 
anisotropic fracture surfaces.
\end{itemize}
All these models lead to self affine fracture 
surfaces, but with varying critical exponents. 
The relationship between these predicted 
exponents and the roughness exponent measured 
experimentally remains controversial.

We present here an extensive experimental 
investigation of the scaling properties of 
fracture surfaces in heterogeneous materials. Four 
different materials with four different 
characteristic length scales of the microstructure 
were investigated: silica glass, aluminum alloy, 
mortar and wood. They were broken using four 
different types of fracture tests and the 
fracture surfaces were scanned by three different 
techniques. Fracture surfaces observed in all 
these different materials/failure modes are shown 
to be self-affine, in agreement with results 
reported in the literature. However, their 
scaling properties are not isotropic as usually 
believed but require the use a two-dimensional 
(2D) height-height correlation function for a 
complete description.

This 2D description involves two independent 
scaling exponents, which are the Hurst exponents 
measured along the crack front direction and 
along the crack propagation direction. Those will 
be shown to correspond respectively to the 
roughness exponent $\zeta$ and the growth 
exponent $\beta$ introduced to describe the 
depinning of  elastic manifolds in a random 
medium (\opencite{Kardar2}; \opencite{Barabasi}). 
To describe fully the scaling properties of the 
experimental fracture surfaces, a third exponent 
- the dynamic exponent $z$ - must be introduced. 
As expected for Family-Vicsek scaling 
\cite{Family}, this last exponent is shown to be 
equal to the ratio of the two others, 
$z=\zeta/\beta$.

These three exponents are found to vary little 
with the nature of the material and the crack 
propagation velocity. They are conjectured to be 
universal. Implications of these scaling 
properties will finally be discussed.

\section{Description of fracture tests and surface scanning}\label{experiments}

Silica glass and aluminum alloy are archetypes 
of brittle and ductile materials respectively 
while mortar and wood are good examples of 
isotropic and anisotropic quasi-brittle materials 
respectively.

\begin{itemize}
\item[(a)] Silica fracture surfaces were obtained 
by applying a DCDC (Double Cleavage Drilled 
Compression) to parallelepipedic
($5 \times 5 \times 25 \un{mm}^3$) samples under 
stress corrosion in mode I (see 
\inlinecite{Prades} for details). After a 
transient dynamic regime, the crack propagates at 
a slow velocity through the specimen under stress 
corrosion. This velocity was measured by imaging 
in real time the crack tip propagation through 
Atomic Force Microscopy (AFM). In the stress 
corrosion regime, the crack growth velocity can 
be controlled  by adjusting the compressive load 
applied to the DCDC specimen \cite{Bonamy}. The 
protocol is then the following: (i) a large load 
is applied to reach a high velocity; (ii) the 
load is decreased to a value lower than the 
prescribed one; (iii) the load is increased again 
up to the value that corresponds to the 
prescribed velocity and kept constant. This 
procedure allows to observe on the post-mortem 
fracture surfaces constant velocity zones 
clearly separated by visible  arrest marks. The 
topography of these fracture surfaces is then 
measured through AFM with in-plane and 
out-of-plane resolutions of the order of 
$5\un{nm}$ and 0.1 $\un{nm}$ respectively. To 
ensure that there is no bias due to the scanning 
direction of the AFM tip, each image is scanned 
in two perpendicular directions and the analyses 
presented hereafter are performed on the two sets 
of images. These images represent a square field of 
$1\times1~\mu\mathrm{m}$ (1024 by 1024 pixels).

\item[(b)] Fracture surfaces of the commercial 
7475 aluminum alloy were obtained from CT 
(compact tension) specimens, first precracked in 
fatigue and then broken under uniaxial mode I 
tension. In the tensile zone, the fracture 
surface has been observed with a scanning 
electron microscope at two tilt angles. A high 
resolution elevation map has been produced from 
the stereo pair using the cross-correlation based 
surface reconstruction technique described in 
\inlinecite{Amman2}. The reconstructed image of 
the topography represents a rectangular field of 
$565\times405~\mu\mathrm{m}$ (512 by 512 pixels). The in-plane and 
out-of-plane resolutions are of the order of 
$2-3~\mu\mathrm{m}$.

\item[(c)] Mortar fracture surfaces were obtained 
by applying four points bending to a notched beam 
leading to a mode I failure. The displacement is 
controlled during the test. The length of the 
beam is 1400 mm and its height and thickness are 
both equal to 140 mm. The topography of the 
fracture surfaces has been recorded using an 
optical profilometer. The maps include 500 
profiles of 4096 points each with the first 
profile close to the initial notch. The sampling 
step along profiles is $20~\mu\mathrm{m}$.  Two 
successive profiles are separated by 
$50~\mu\mathrm{m}$ along the direction of crack 
propagation. The lateral and vertical accuracy 
are of the order of $5~\mu\mathrm{m}$. A 
transient regime is also observed by a 
post-mortem analysis of the fracture surfaces. On 
the first 10$\un{mm}$ of the crack  propagation, 
corresponding to the first 200 profiles, the 
roughness of the profiles increases with the 
distance to the initial straight notch. A full 
description of the roughness in this region of 
the fracture surface is given in 
\inlinecite{Mourot2}. The present study focuses 
on the geometry of the surface  far from the 
initial notch and the first 200 profiles are 
therefore systematically removed from the maps.

\item[(d)] Fractured wood surfaces were obtained 
from modified Tapered Double Cantilever Beam 
specimens (TDCB) subject to uniaxial tension with 
a constant opening rate leading to mode I failure 
(see \inlinecite{Morel6} for details). The wood 
species used in the study is a Spruce ({\it Picea 
excelsa W.}) which is strongly anisotropic. The 
crack propagated along the longitudinal direction 
of the wood. As a result, the characteristic 
length scale of the elementary feature of the fracture surface is 
anisotropic : it is respectively of the order of 
a mm and of a few tenths of micrometer in the longitudinal and transverse directions. These 
values correspond respectively to the length and 
the diameter of the wood cells. As a consequence, 
the height of the surface has been scanned by an 
optical profilometer  over a $50\times 50\un mm$ 
area with a higher resolution in the transverse 
direction ($25~\mu\mathrm{m}$) than in the 
longitudinal one ($2.5\un{mm}$) : this map 
includes  50 profiles parallel to the crack front 
with  2048 points each. As for the mortar 
fracture surfaces, the maps of the wood fracture 
surfaces correspond to the zone far from the 
initial straight notch where the roughness is 
statistically stationary, {\it i. e.} 
approximately 50$\un{mm}$  from initiation.
\end{itemize}

\begin{figure}[!h]
\includegraphics[width=0.49\columnwidth]{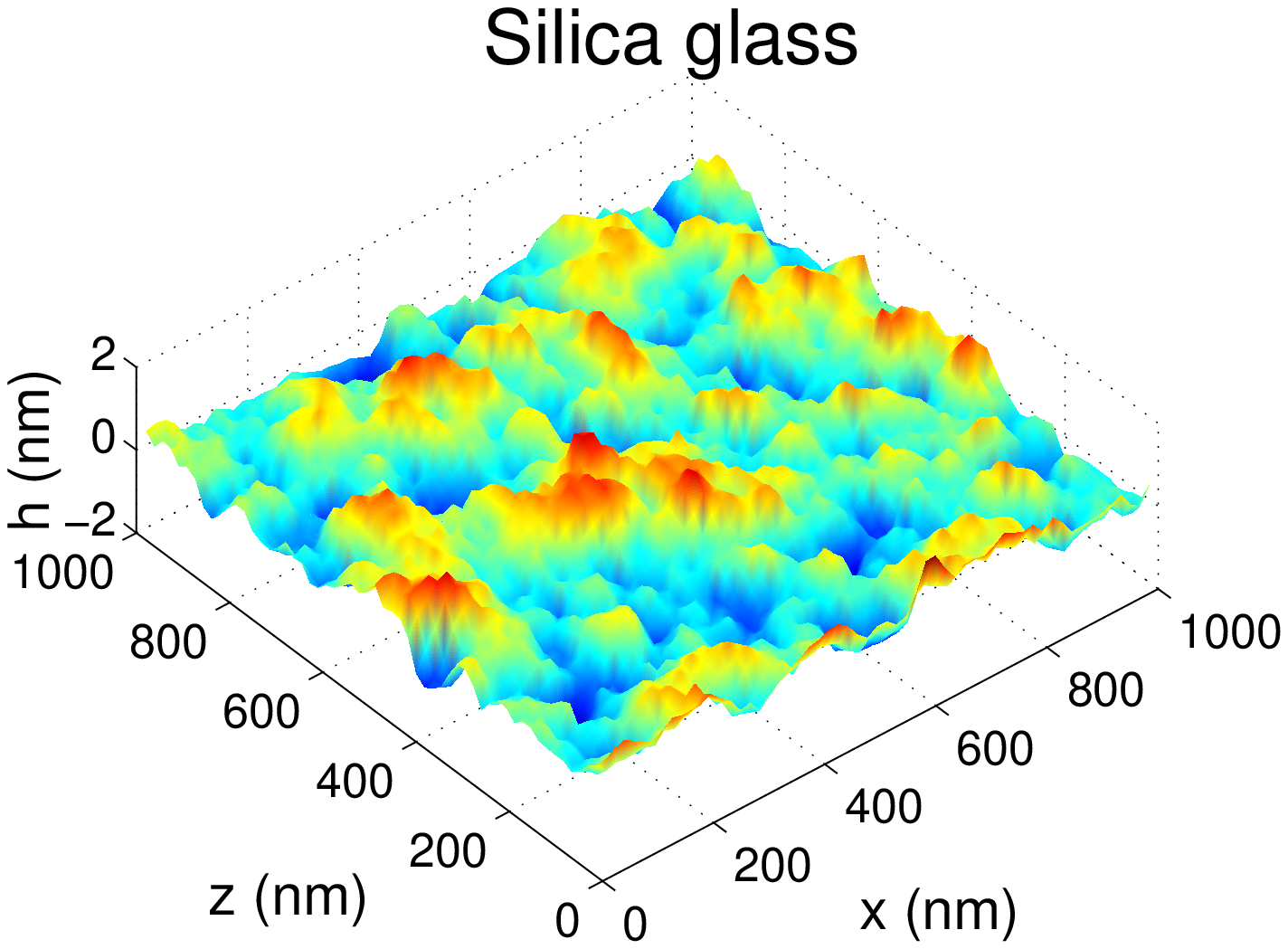}
\includegraphics[width=0.49\columnwidth]{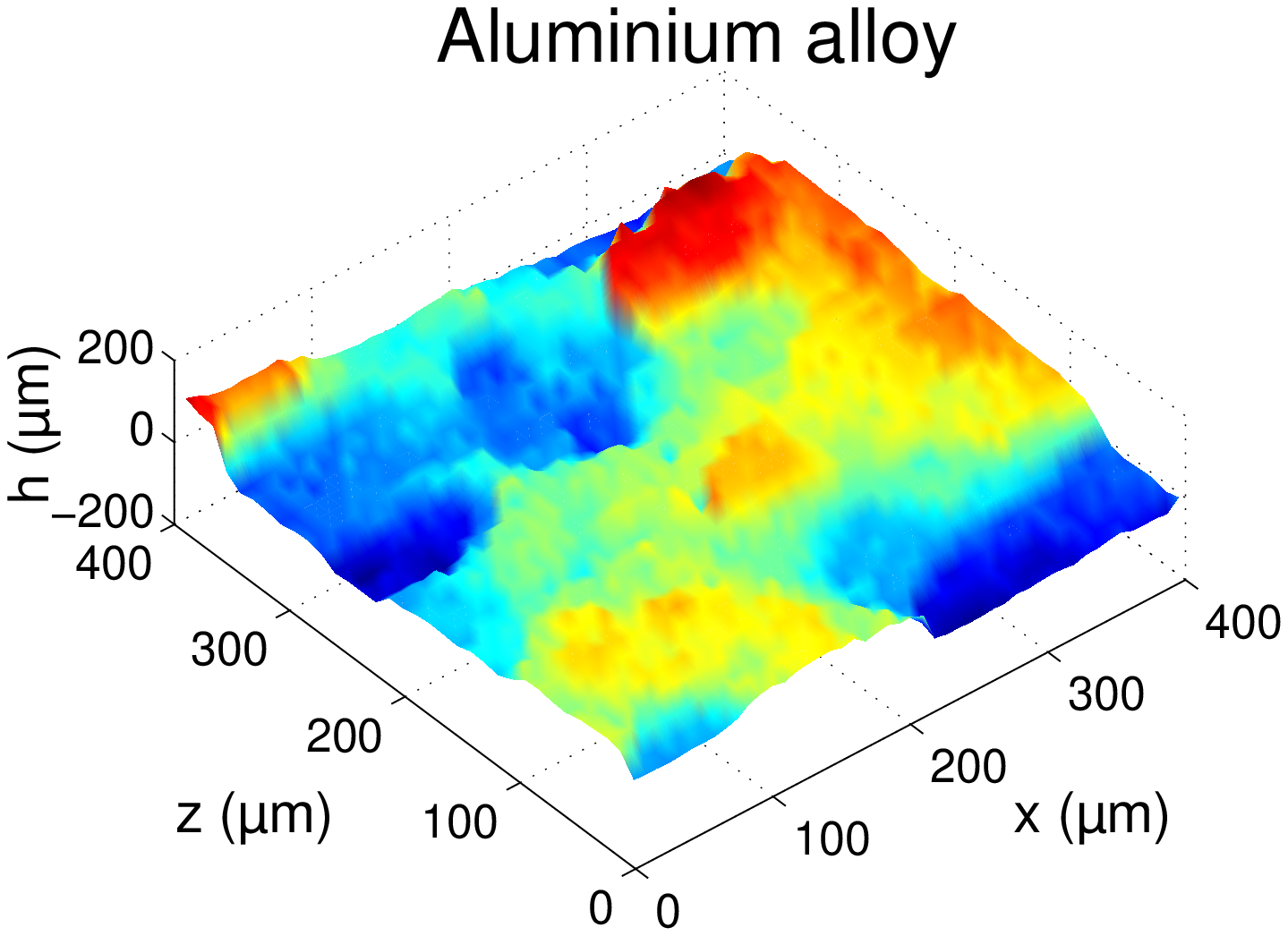}
\includegraphics[width=0.49\columnwidth]{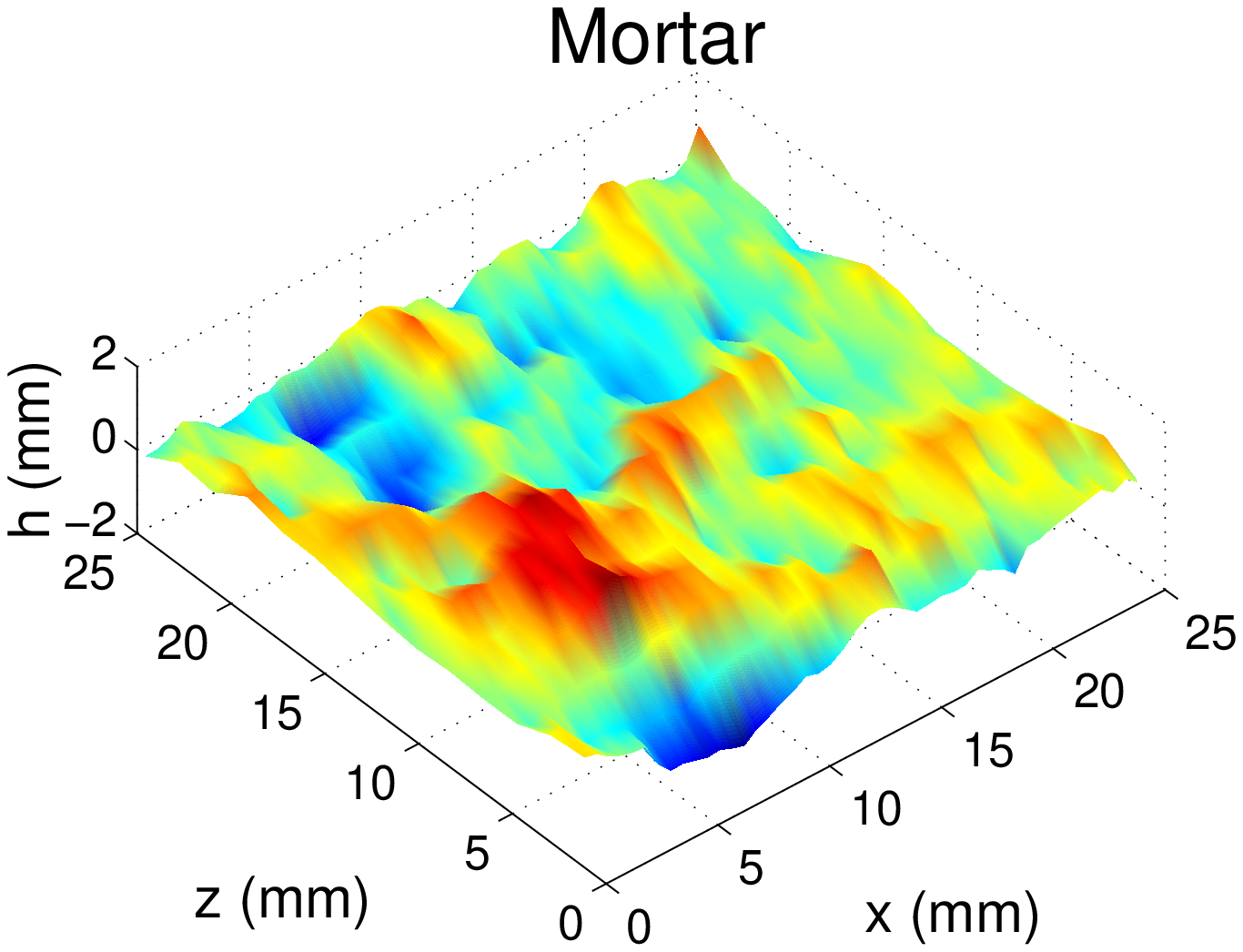}
\includegraphics[width=0.49\columnwidth]{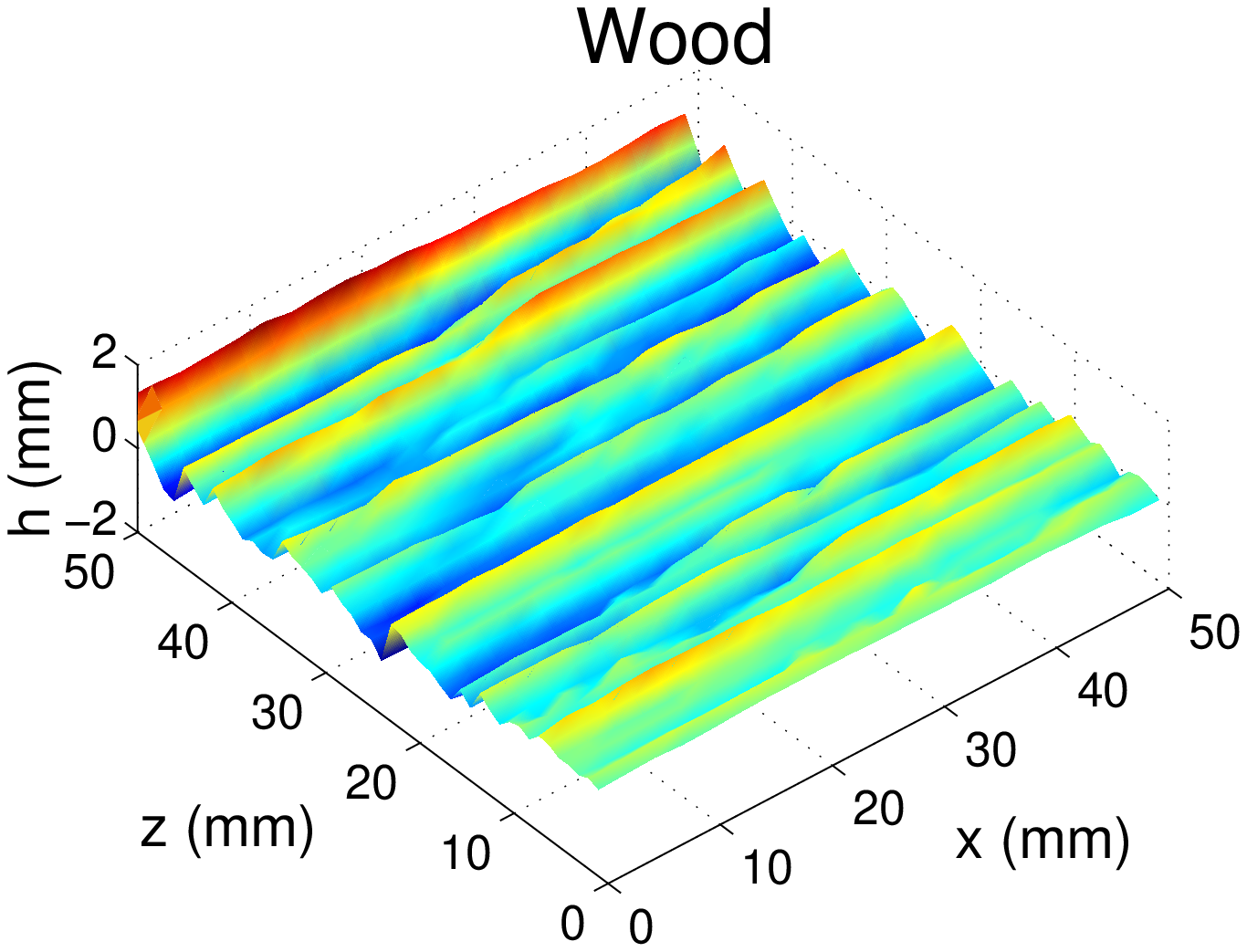}
\centering
\caption{Topographic images of fracture surfaces 
of silica glass, aluminum alloy, mortar and wood. 
Square fracture surfaces were represented here 
for the sake of clarity. The $x$-axis and 
$z$-axis correspond to the crack propagation 
direction and the crack front direction 
respectively.}
\label{surfaces}
\end{figure}

In all four cases, the reference frame 
$(\vec{e}_x,\vec{e}_y,\vec{e}_z)$ is chosen so 
that $\vec{e}_x$ and $\vec{e}_z$ are respectively 
parallel to the direction of crack propagation 
and to the crack front. Figure \ref{surfaces} 
shows typical snapshots of the fracture surfaces 
as observed in silica glass, aluminum alloy, 
mortar and wood. These surfaces display striking 
visual differences : the in-plane (along $x$ or 
$z$) and out-of-plane (along $h$) length-scales 
of the features observed strongly depend indeed 
on the considered material. They are respectively 
of the order of  50$\un{nm}$ and 1$\un{nm}$ for 
the silica glass surface, approximately 
$100~\mu\mathrm{m}$ and $30~\mu\mathrm{m}$ for 
aluminum, and $5$ and 0.6$\un{mm}$ for mortar. 
The fracture surface in wood is highly 
anisotropic: The in-plane sizes of the patterns 
are respectively $50$ and 1$\un{mm}$ along the 
longitudinal ($x$-axis) and transverse ($z$-axis) 
directions and  $200~\mu\mathrm{m}$ for out of 
plane sizes.

Despite their apparent differences, these 
surfaces share common scaling properties to be 
discussed in the next section.

\section{Fracture surface analysis: Anisotropic 
scaling properties}\label{anisotropy}

\begin{figure}[!h]
\includegraphics[width=0.55\columnwidth]{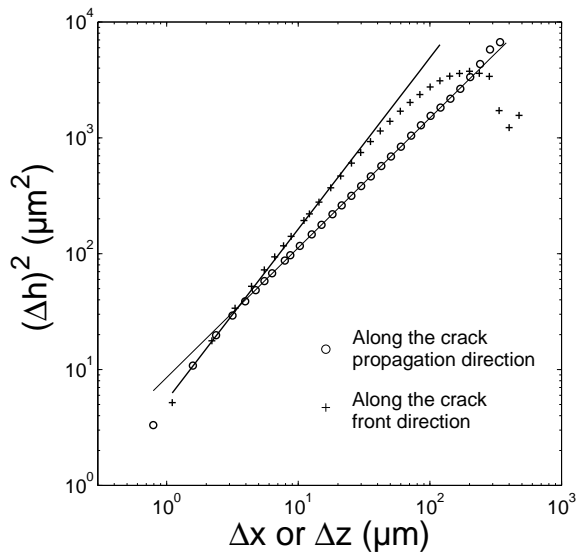}
\centering
\caption{1D height-height correlation functions 
measured parallel to the crack propagation 
direction and to the crack front for an aluminum 
alloy fracture surface. The straight lines are 
power law fits. The scaling exponents, corresponding to the slope of these lines,
are respectively 
equal to  0.75 and 0.58 parallel to  the crack 
front  and to the crack propagation.}
\label{anisotropy_fig}
\end{figure}

In order to investigate the anisotropy of the 
experimental fracture surfaces, the 1D 
height-height correlation functions $\Delta 
h(\Delta z)=<(h(z+\Delta z,x)-h(z,x))^2>^{1/2}$ 
along the $z$ direction, and $\Delta h(\Delta 
x)=<(h(z,x+\Delta x)-h(z,x))^2>^{1/2}$ along the 
$x$ direction were computed for each material. 
They are represented in Figure 
\ref{anisotropy_fig} for a fractured aluminum 
alloy sample.

These curves indicate a clear dependence on the 
measurement direction although all profiles are 
self affine. This anisotropy is reflected not 
only in the correlation lengths and in the 
amplitudes, {\em but also in the values of the 
Hurst exponent}: Along the crack front, it is 
found to be $\zeta = 0.75 \pm 0.03$, {\it i. e.} 
fairly consistent with the "universal" value of 
the roughness exponent $\zeta \simeq 0.8$ 
reported in the literature (\opencite{Bouchaud9}; 
\opencite{Maloy}; \opencite{Schmittbuhl8}; 
\opencite{Daguier}). Parallel to the crack front, 
the measured Hurst exponent is significantly 
smaller, with $\beta=0.58 \pm 0.03$.

The observation  of two different scaling 
behaviors in two different directions of the 
fractured aluminum alloy surface suggests a new 
approach  based on the analysis of the 2D 
height-height correlation function defined as:

\begin{equation}
\Delta h(\Delta z,\Delta x)=<[h(z+\Delta z,x+\Delta x)-h(x,z)]^2>^{1/2}_{z,x}
\end{equation}

\begin{figure}
\centering
\includegraphics[height=0.35\columnwidth]{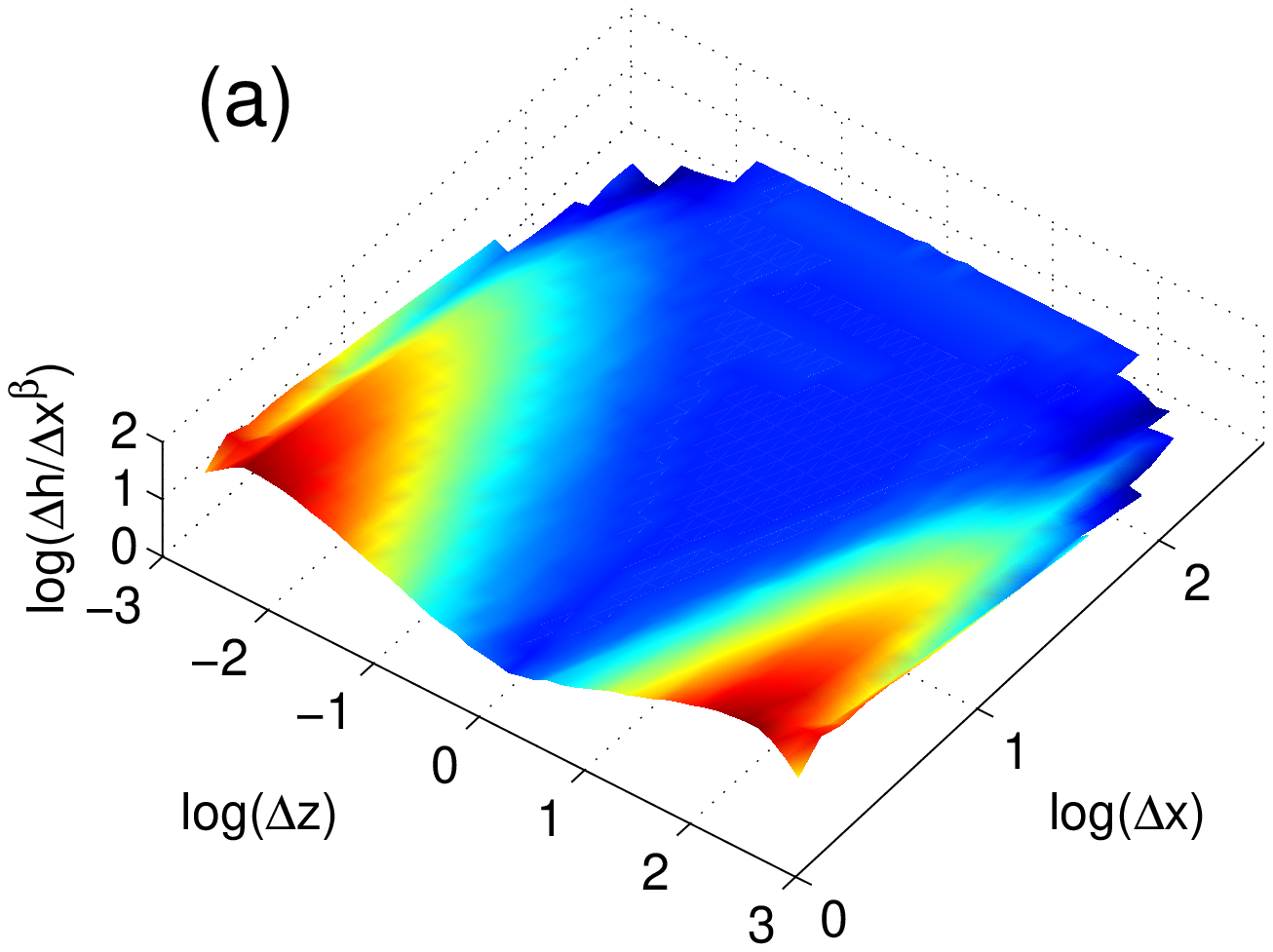}
\includegraphics[height=0.35\columnwidth]{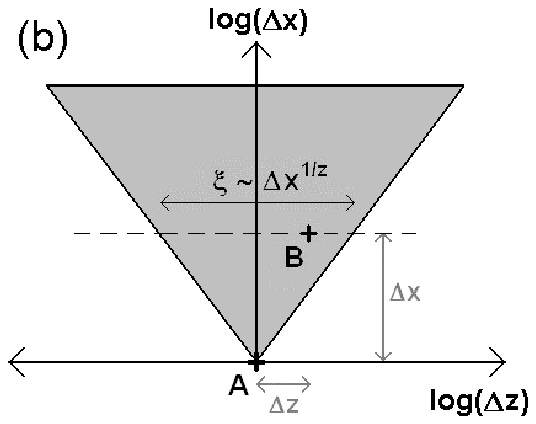}
\caption{(a):  2D representation of the 
height-height correlation function for an 
aluminum fracture surface ($\Delta h$ is 
normalized by $\Delta x^\beta$ with 
$\beta=0.58$). (b): Domains of different types of 
variation of the height-height correlation 
function  in the $(\Delta z, \Delta x)$ plane. 
The grey zone corresponds to a height-height 
correlation function varying as $\Delta h \propto 
\Delta x^\beta$}
\label{cone}
\end{figure}

\noindent This function contains informations on 
the scaling properties of a surface in {\em all} 
directions. Figure \ref{cone}a gives a color 
scale representation of $\Delta h$ in the ($\Delta 
z,\Delta x$) plane  for the aluminum alloy. The 
function $\Delta h$ was normalized by $\Delta 
x^\beta$ and logarithmic scales were used for all 
the axis  to emphasize the anisotropy of the 
power-law scalings. This representation clearly 
demonstrates two distinct behaviors of the 2D 
correlation function depending on the orientation 
of the vector $\vec{AB}$ of coordinates $(\Delta 
z,\Delta x)$.

If $\vec{AB}$ lies inside the grey region  in 
Figure \ref{cone}b (corresponding to the blue 
domain in Figure \ref{cone}a), the 2D correlation 
function scales as $\Delta x^\beta$ and does not 
depend on $\Delta z.$ The straight boundaries of 
this domain in these logarithmic scales indicate 
that its width $\xi$ (Figure \ref{cone}b) 
increases following a power law  $\xi \propto \Delta 
x^{1/z}$ where $z \simeq 1.26.$

In other words, from any given point A of the 
fracture surface, a domain where the 2D 
correlation function scales as $\Delta x^\beta$ 
develops over a width  $\Delta z = \xi$ 
increasing as $\Delta x^{1/z}$ (the crack 
propagates parallel to $x$). Outside of this 
domain, the 2D correlation function depends only 
on $\Delta z$. If $\Delta x$ =0, 
the correlation function corresponds to that 
presented in Figure \ref{anisotropy_fig} and 
scales as $\Delta h \propto \Delta z^{\zeta}$ 
with $\zeta=0.75$. The latter variation  is not 
visible in Figure \ref{cone}a because of the 
divergence of the normalization term $\Delta 
x^{\beta}.$

These scaling behaviors of the 2D 
correlation function of the aluminum fracture 
surface can be summed up as follows:

\begin{equation}
\begin{array} {l}
    \Delta h(\Delta z,\Delta x)=\Delta x^{\beta}f(\Delta z/\Delta x^{1/z}) \\
\\
where\quad	f(u) \sim \left\{
\begin{array}{l l}
1 & $if u$ \ll c  \\
u^{\zeta} & $if u$ \gg c

\end{array}
\right.
\end{array}
\label{cor2D_eq}
\end{equation}

\noindent where the constant $c$ is related to 
the topothesies $l_x$ and $l_z$ defined along $x$and 
$z$ respectively.

\begin{figure}
\centering
\includegraphics[width=0.49\columnwidth]{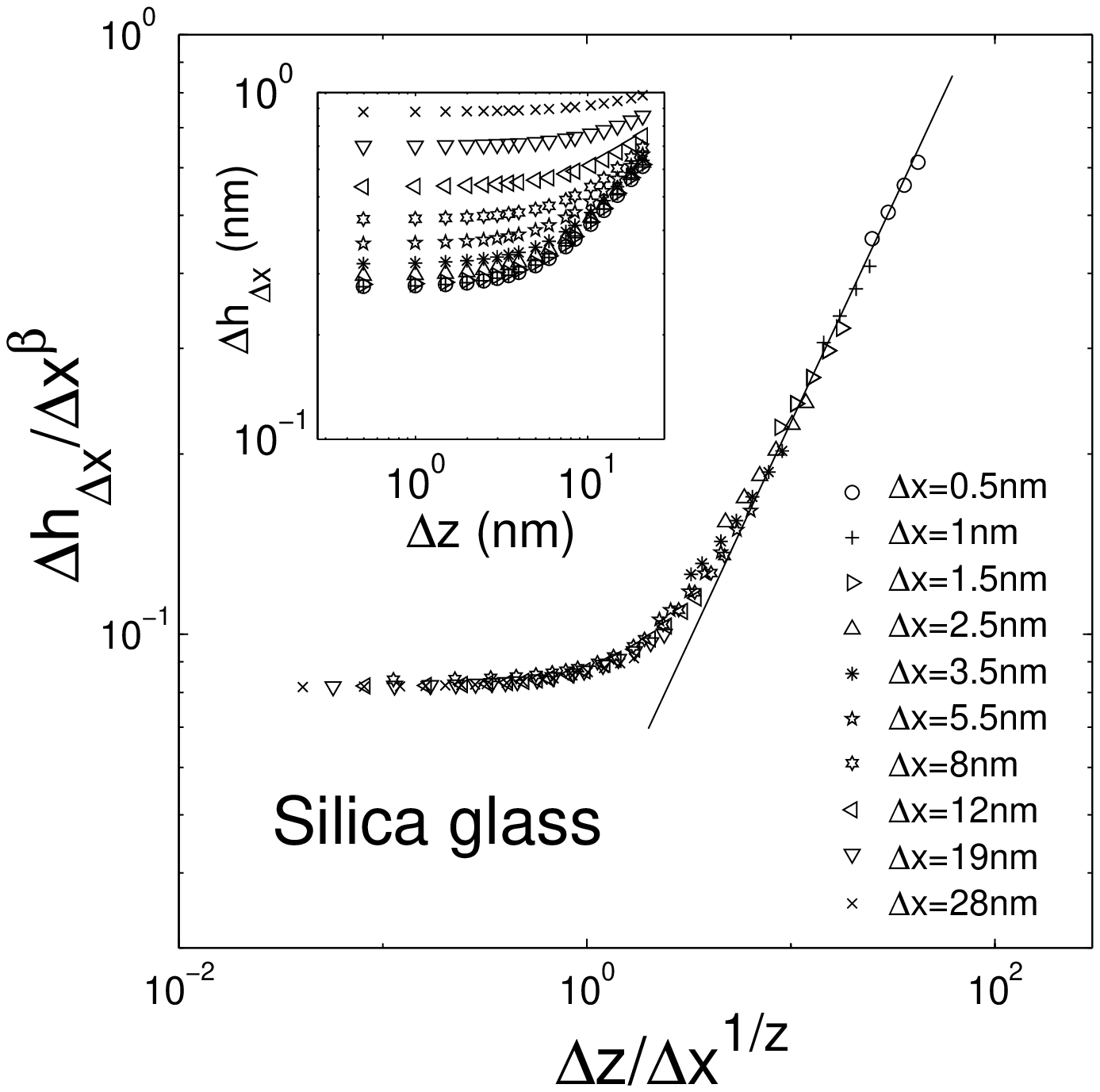}
\includegraphics[width=0.49\columnwidth]{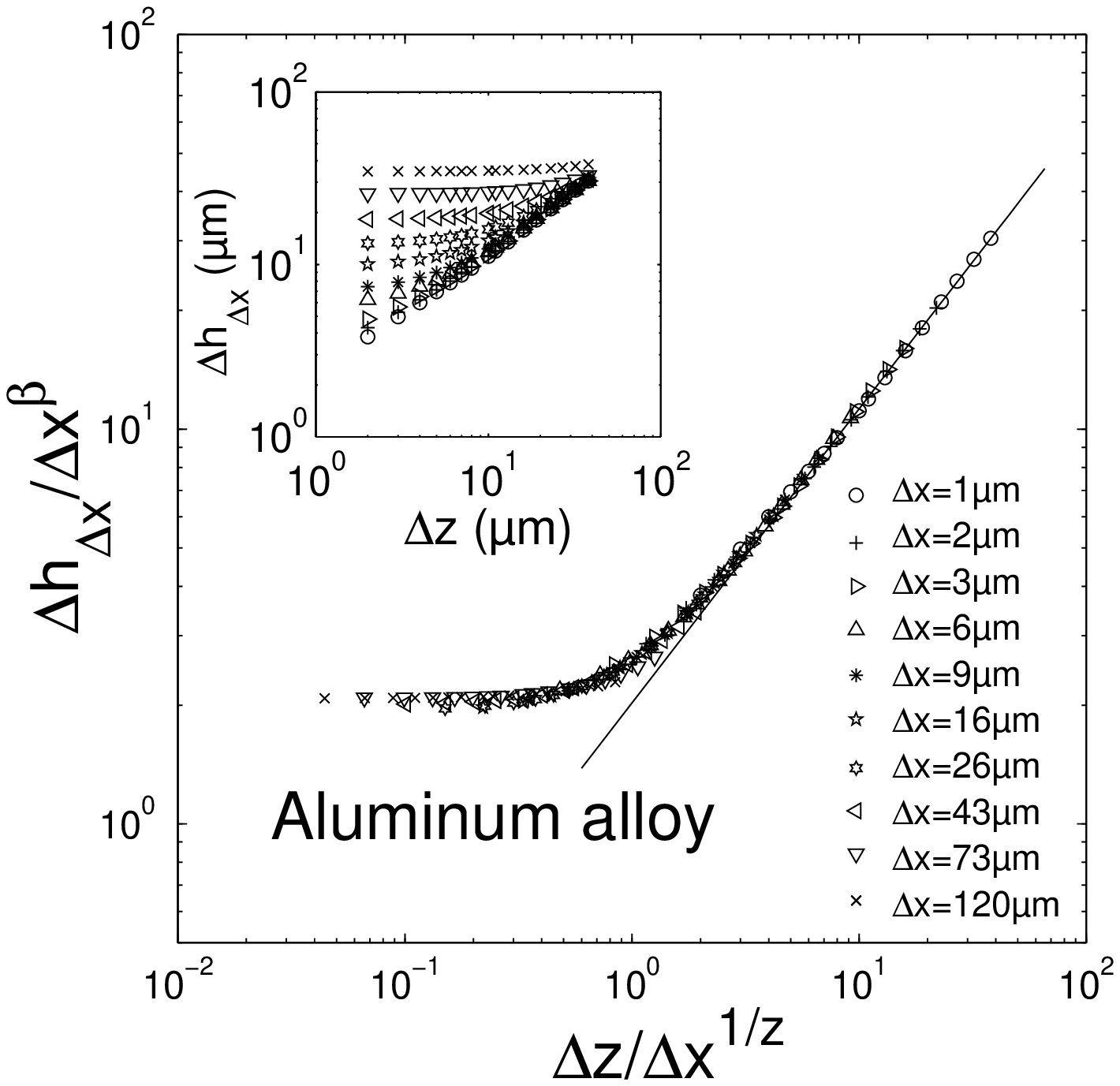}
\includegraphics[width=0.49\columnwidth]{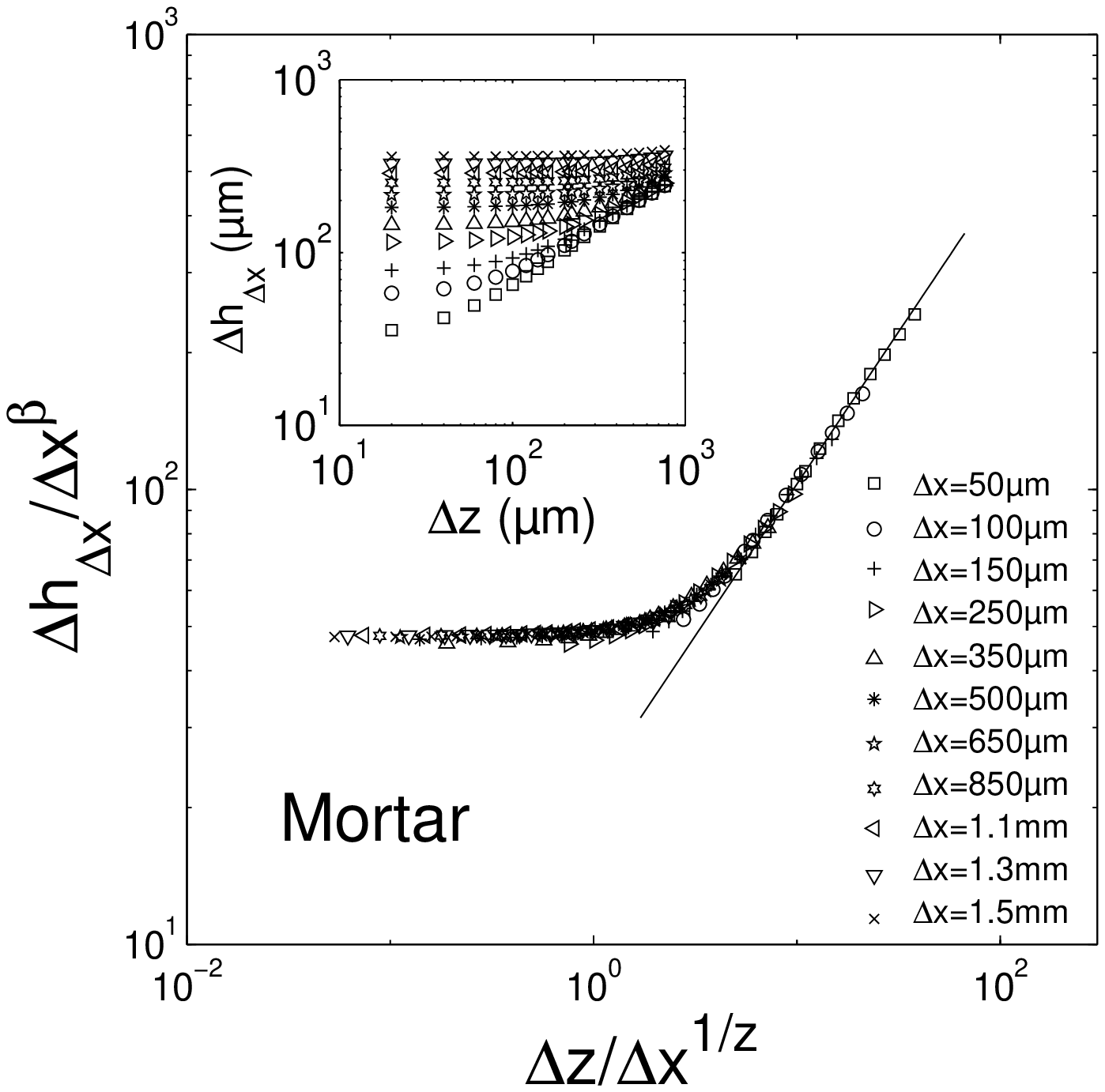}
\includegraphics[width=0.49\columnwidth]{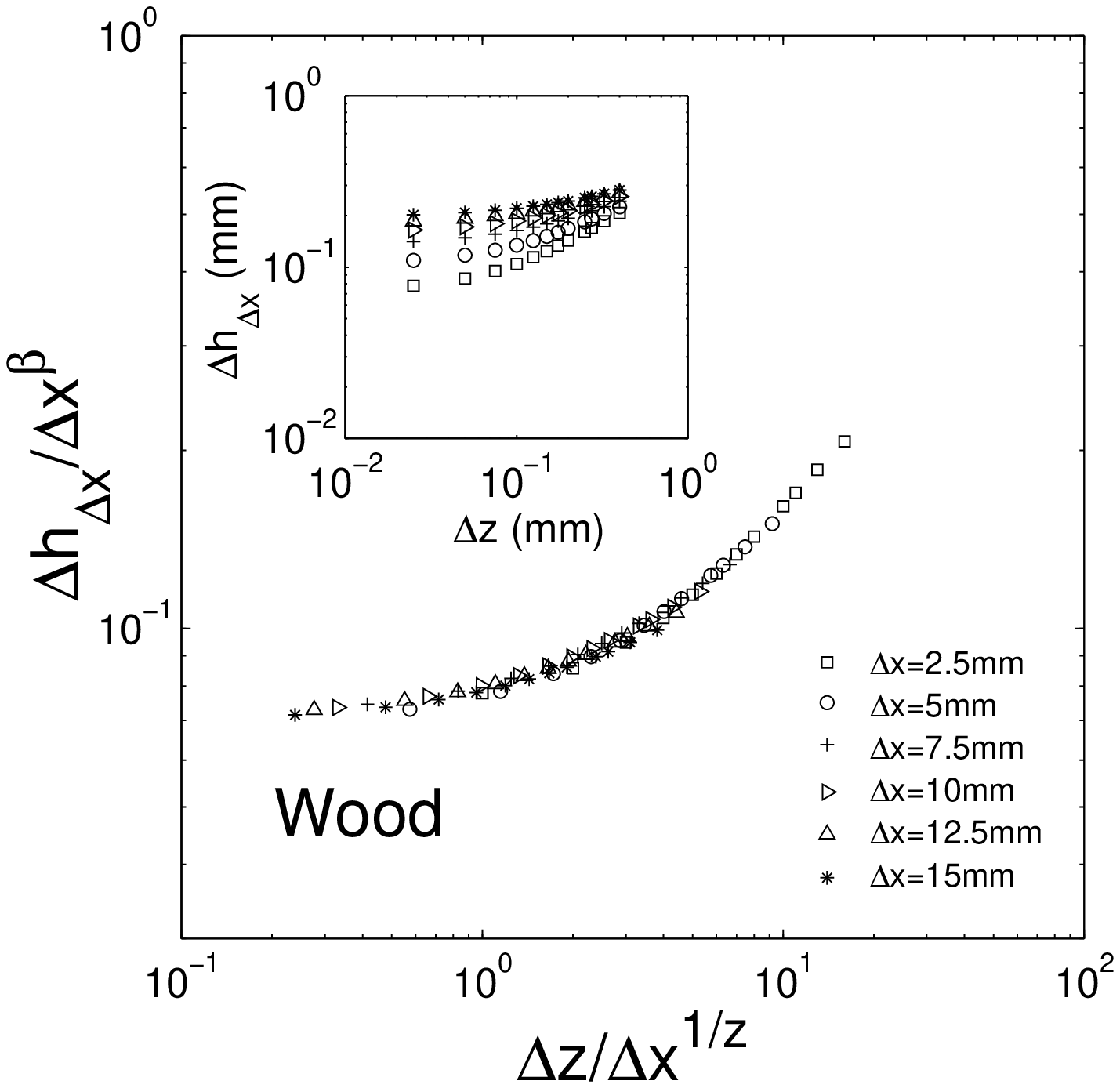}
\caption{Normalized 2D height-height correlation 
function variations with  $\Delta z$ for various 
values
of $\Delta x$ for silica glass, aluminum alloy, 
mortar and wood. The data collapse was obtained 
from Eq. \ref{cor2D_eq} using exponents
reported in Tab. \ref{tab1}.}
\label{cor2D_fig}
\end{figure}

The variations of the correlation functions 
$\Delta h_{\Delta x}$ are plotted as a function 
of $\Delta z$ in the insets of Figure 
\ref{cor2D_fig} respectively for silica glass, 
aluminum alloy, mortar and wood fracture 
surfaces. This Figure demontrates that the 
scaling properties described by Equation 
\ref{cor2D_eq} apply for all fracture surfaces investigated
experimentally. For adequate values 
of $\beta$ and $z$, it can be seen in the main 
graphs of that same figure that a very good 
collapse of the curves can be obtained by 
normalizing the x-axis by $\Delta x^{1/z}$ and 
the y-axis by $\Delta x^{\beta}$. As expected 
from Equation \ref{cor2D_eq}, the resulting 
master curve is characterized by a plateau region 
and followed  by a power law variation of
exponent $\zeta$. The exponents $\beta$ and $z$ 
which optimize the collapse, and the $\zeta$ 
exponent determined by fitting the large scales 
regime exhibited by the master curves are listed 
in Table \ref{tab1}. The three exponents are 
found to be $\zeta \simeq 0.75$, $\beta \simeq 
0.6$ and $z \simeq 1.25$, independent of the 
material and of the crack growth velocity over 
the whole range from ultra-slow stress corrosion 
fracture (picometers per second) to rapid
failure (some meters per second). They can 
therefore be conjectured to be {\em universal}.

The ratio of $\zeta$ to $\beta$ is given in the 
fourth column. It is worth noting that the 
exponent $z$ verifies the relation 
$z=\beta/\zeta$ as expected for Family-Vicsek 
scalings (\opencite{Family}) classically observed 
in interface growth processes.

\begin{table}[!h]
\centering
\begin{tabular}{|>{\centering}m{70pt}
                 |>{\centering}m{85pt}
                 |>{\centering}m{85pt}
                 |>{\centering}m{85pt}
                 |c|}
\hline
  & $\zeta$ & $\beta$ & z & $\zeta / \beta$\\
\hline
silica glass & 0.77 $\pm$ 0.03 & 0.61 $\pm$ 0.04 & 1.30 $\pm$ 0.15 & 1.26 \\
aluminum alloy & 0.75 $\pm$ 0.03 & 0.58 $\pm$ 0.03 & 1.26 $\pm$ 0.07 & 1.29\\
mortar & 0.71 $\pm$ 0.06 & 0.59 $\pm$ 0.06 & 1.18 $\pm$ 0.15 & 1.20\\

wood & 0.79 $\pm$ 0.05 & 0.58 $\pm$ 0.05 & 1.25 $\pm$ 0.15 & 1.36\\
\hline
\end{tabular}
\caption{Scaling exponents measured on fracture 
surfaces of silica glass, metallic alloy, mortar 
and wood. $\zeta$, $\beta$, z and $\zeta/ \beta$ 
are respectively the roughness exponent,  the 
growth exponent and  the dynamic exponent $z$ 
while the fourth column contains the ratio of 
$\zeta$ to $\beta$. Error bars correspond to a 
confidence interval  of 95 \%.}
\label{tab1}
\end{table}

\section{Discussion}\label{discussion}

Equation \ref{cor2D_eq} suggests analogies with 
roughening in interface growth processes. In Ref. 
\inlinecite{Ponson5}, surfaces were generated 
numerically by solving the Edwards-Wilkinson 
(\opencite{Edwards}) and the Kardar-Parisi-Zhang 
(\opencite{Kardar}) equations, which describe the 
dynamics of an elastic line perturbed by a 
thermal noise. In such systems, the roughness 
developed from a flat profile in the transient 
initial regime is known to be characterized by a 
1D height-height cross-correlation function 
$\Delta h(\Delta z,t)=<(h(z+\Delta 
z,t)-h(z,t))^2>_x^{1/2}$ with a time dependence 
verifying \cite{Barabasi}:

\begin{equation}
\begin{array} {l}
    \Delta h(\Delta z,t)= t^ {\beta^*} g(\Delta z/t^{1/z^*}) \\
\\
where\quad	g(u) \sim \left\{
\begin{array}{l l}
u^{\zeta^*} & $if u$ \ll d  \\
1 & $if u$ \gg d
\end{array}
\right.
\end{array}
\end{equation}

\noindent where $d$ is constant and $\zeta^*$, 
$\beta^*$ and $z^*$ refer to the roughness, 
growth and dynamic exponents respectively, as 
commonly defined in the framework of elastic 
manifolds propagating in a random environment. 
These three exponents are not independent and are 
related by $z^*=\zeta^*/\beta^*$ 
(\opencite{Family}).

In Ref.\inlinecite{Ponson5},  the 2D 
height-height correlation function $\Delta 
h(\Delta z,\Delta t)$ is computed for these 
simulated surfaces in the steady state regime 
reached at long times where the roughness becomes 
time invariant. It is shown that $\Delta h(\Delta 
z,\Delta t)$ verifies the scaling Equation 
\ref{cor2D_eq} where the exponents $\zeta$, 
$\beta$ and $z$ coincide with $\zeta^*$, 
$\beta^*$ and $z^*$ respectively. In other words, 
the "universal" exponents $\zeta \simeq 0.75$, 
$\beta \simeq 0.6$ and $\beta \simeq 1.25$ 
measured on experimental fracture surfaces in the 
previous section correspond respectively to the 
roughness, growth and dynamic exponents.

Note that these results differ from those of 

previous measurements in the initial transient 
regime   which reported material-dependent 
dynamic exponents (\opencite{Schmittbuhl2}; 
\opencite{Lopez4}; \opencite{Morel6}; 
\opencite{Mourot2}). These two last studies 
focused on the fracture surface morphology, 
respectively of mortar and wood,  in the 
roughening regime starting from a straight notch. 
The surfaces were found to be anisotropic and to 
display an anomalous scaling requiring to 
introduce a fourth, global roughness, exponent. 

Moreover, the dynamic exponents were found to be 
different from those reported in the present 
study on the same materials and to vary from material 
to material. In other words, 
there is apparently no relationship between the 
exponents measured in the transient and steady 
state propagation regimes.

This observation is quite surprising in view of 
the predictions of theoretical models. In models 
based on elastic lines propagating in random 
environments, the physical mechanism which rules 
the time evolution of the line is not changing 
between the transient and the long-time regime. 
Hence, the exponents are found the same in the 
two regimes \cite{Ponson5}. In experiments, we 
would thus expect to get also the same exponents in 
transient and stationary regimes. However, in 
quasi-brittle materials, the physical mechanisms 
that rule the roughening of a fracture surface 
developing from an initial straight notch 
(microcracks development, material toughness,...) 
is evolving with time during the transient 
regime. This may explain the apparent discrepancy 
in measured exponents.

\section{Conclusion}

The scaling properties of fracture surfaces of 
materials with different characteristic length 
scales of their microstructure have been 
investigated. Measurements  were realized on 
brittle (amorphous silica), ductile (aluminum 
alloy) and quasi-brittle (mortar and wood) 
materials. The surfaces were studied far enough 
from the initiation of the crack so that  the 
roughness distribution  could be considered as 
stationary.  It was first observed on aluminum 
alloy surfaces that profiles parallel and 
perpendicular to the direction of crack 
propagation were both self-affine, but 
corresponded to different Hurst exponents. In 
order to characterize such surfaces, 1D 
measurements such as roughness profiles are too 
restrictive, and the 2D height-height correlation 
function has been shown to be more suitable.
For all types of materials investigated, the 2D 
correlation function displays similar scaling 
properties involving three exponents $\zeta 
\simeq 0.75$, $\beta \simeq 0.6$ and $z \simeq 
1.25$ independent of the considered material as 
well as the crack growth velocity. They are shown 
to correspond respectively to the roughness, 
growth and dynamic exponents introduced  in 
interface growth models. Moreover, the three 
exponents are not independent: $z=\zeta/\beta$ as 
expected for Family-Viseck scaling 
(\opencite{Family}).

The fact that $z \not= 1$ and $\beta \not= \zeta$ 
shows that scaling features of fracture surfaces 
result from {\em dynamic} processes. In others 
words, static models like percolation-based 
models \cite{Hansen} are not suitable to describe 
fracture surfaces in heterogeneous materials. The 
experimental observations reported here provide 
strong arguments in favor of models based on 
elastic manifolds moving in a random medium. As 
shown by \inlinecite{Roux} and 
\inlinecite{Rosso3} in simulations of the 
propagation of an elastic line in a random 
environment, the criticality of such a phenomenon 
is expected to arise only at long times. This is 
in perfect agreement with the experimental 
evidence of the existence of three universal 
exponents in the long time regime reported here. 
It should be emphasized that elastic strings 
models available presently don't succeed in reproducing 
quantitatively the exponents observed. Work in 
this direction is currently under progress.

\begin{acknowledgements}
We are indebted to J. P. Bouchaud and S. Roux for enlightening discussions.
\end{acknowledgements}

\bibliographystyle{klunamed}

\end{article}

\end{document}